\documentclass[sigconf,nonacm]{acmart}
\renewcommand\footnotetextcopyrightpermission[1]{}
\begin{document}

\title{AI Agent Pull Requests on GitHub: Frequency, Structure, and Merge Conflict Rates}

\author{George Xu}
\authornote{All authors contributed equally to this work.}
\affiliation{%
  \institution{Harvard Medical School / Massachusetts General Hospital}
  \city{Boston}\state{Massachusetts}\country{USA}}

\author{Arjun Subramanian}
\authornotemark[1]
\affiliation{%
  \institution{Massachusetts Institute of Technology Computer Science \& Artificial Intelligence Lab}
  \city{Cambridge}\state{Massachusetts}\country{USA}}

\author{Nithilan Karthik}
\authornotemark[1]
\affiliation{%
  \institution{DevRev AI LLC}
  \city{San Francisco}\state{California}\country{USA}}
  
\begin{abstract}
AI coding agents may generate and submit Pull Requests (PRs) to the same repository at the same time. However, research concerning the extent of concurrent submission by AI coding agents to a common repository does not exist. This paper uses the AIDev-pop dataset (33,596 PRs in 2,807 repositories) to provide the first empirical examination of the prevalence of concurrent submission using PRs authored by agents. We report that when considering exact temporal overlap, 40.2\% of repositories contain co-active agent-authored PR pairs; further, the co-active pairs account for 79.4\% of all PRs generated by an AI agent. When we examine co-activity within a one week collaboration window, the percentages are increased to 53.4\% and 95.0\%, respectively. For the majority of the co-active PR pairs (underlying the vast majority of which are intra-agent authored), both PRs were authored by the same agent, while only 0.5\% of co-active pairs were cross-agent, and occurred in only 122 out of 2807 total repositories examined (or approximately 4.3\%). Additionally, we replayed actual three way git merges on 747 unique co-active pairs (one per repository), and computed the percentage of textual conflict encountered during the merge operation to combine the two PRs in each pair. We observed that the percentage of textual conflict encountered was significantly higher for cross-agent pairs compared to intra-agent pairs: 41.7\% vs.\ 19.8\%, respectively, with non-overlapping 95\% confidence intervals. Lastly, we developed a classification system based on the detection of conflict reported by git, and determined that the majority of conflicts resulted from modifications to source code files (84.4\% of conflicted files) and not dependency manifest files; further, nearly 42\% of conflicts we observed were structural (i.e., modify/delete or add/add). Since these metrics are limited to the textual level of granularity, they represent conservative lower bounds on costs incurred due to the uncoordinated contributions of multiple AI teammates contributing to a single project.
\end{abstract}


\maketitle

\section{Introduction}
Engineering automation has undergone considerable progress since the first generation of linear, user-interaction-based auto-completion utilities. Today's engineering agents operate independently and in real-time; they have complete environmental understanding and therefore can independently assess issues, write fixes for applications, conduct internal testing cycles, and submit pull requests (PRs) to the operational master branch [1] autonomously. Over the past few years, datasets such as the AIDev ecosystem data set [2], which includes an empirical collection of more than 932,000 pull requests submitted across 116,211 multi-scaled GitHub repositories, have demonstrated that automated developers are creating source code in unprecedented numbers, which indicates a shift toward hybrid or completely autonomous collaborative programming.

A structural issue of coordination between architecture and text arises due to the increased use of autonomous development of source code. In addition to having numerous agent instances---or various agent types---deployed to a single target project at the same time, the typical outcome is a large degree of overlap during each instance's work. Even though standard branching models for git-based systems support two separate PRs targeting the same base commit being built, tested, and run independently after both are merged into a common destination branch, they can immediately cause failures in the application. While traditional version control tools (e.g., git) can identify conflicts between developers based upon specific lines of text when merging changesets, they cannot determine conflicting intent among multiple developers.

The current literature contains much information concerning merge conflicts caused by human activity. This includes characterizing patterns of behavior based on social, structural, and evolutionary metrics across the globe in terms of open-source communities [8][9]. Furthermore, there is a growing body of evidence regarding individual coding agents updating repositories and how successful their updates are in peer review [6][7]. Recent studies have also quantitatively measured how frequently a single PR submitted by an agentic developer will conflict with its base branch [3]. However, there is very little known about the implications for structure of multiple agents working together on a single software project and developing PRs that will conflict with one another at approximately the same time---rather than conflicts between a single PR and its base. An empirical gap exists with respect to two critical variables: (i) the frequency at which concurrently developed agentic workflows exist on GitHub, (ii) and the overall likelihood that these workflows that are developed simultaneously will conflict with each other.

To address this gap, our study provides an empirical examination of the AIDev-pop dataset. From this dataset, we found that there were 2,807 active repositories where 33,596 pull requests were created by authors who were automated developers. Unlike previous research that used heuristically proxied methods to estimate conflict rates, we developed a fully automatic process that could simulate all merge operations using headless git merge tree functionality so as to obtain accurate measurements of actual conflict rates. Our study examines four foundational research questions:
\begin{enumerate}
\item \textbf{RQ1 (Frequency):} For a given time horizon(s), how frequent do agentic PRs become temporally coactive (concurrently active) within a single repository?
\item \textbf{RQ2 (Composition):} What percentage of co-active PR pairings occur from identical intra-agents vs.\ dissimilar cross-agents deployment scenarios?
\item \textbf{RQ3 (Collision):} How frequently do collisions occur when two or more concurrently operating agentic PRs are programmatically combined and merged into the main branch? What characteristics distinguish these conflicts structurally?
\item \textbf{RQ4 (Association):} Which programming languages, sizes of repositories, and types of agents show association(s) with high levels of concurrency?
\end{enumerate}

\section{Background: The Three Layers of Conflict}
To determine how much integration friction will exist from autonomous agents we have identified three structural levels of conflict that arise from taking the two independent modifications ($P_A$ \& $P_B$) and combining them into a single base branch ($B_{\text{base}}$). These conflicts can be divided into three sequential, dependency layers.

\textbf{Conflict Layer 1. Textual Conflicts.} The first layer of conflict arises either through the overlap of modifications made by $P_A$ and $P_B$ to lines of code contained in the same source file; or through a modification that deletes an entire block of code, or an entire file that has also been modified by the other agent. When this type of conflict occurs, the delta-matching engine cannot find a determinate merge path for the two modifications. In these cases, the merge-engine raises a flag indicating a failed merge, generates conflict-markers to the source-file, and stops processing further.

\textbf{Conflict Layer 2. Build Conflicts.} The second level of conflict is produced when $P_A$ and $P_B$ pass through the textual-merging process without raising a failure-flag indicating a failed merge, but upon passing to the compilation stage fail to successfully compile, or fail static-analysis. Typically, this occurs with statically-typed programming languages. For example, $P_A$ modifies the definition of a global function to require an additional input-parameter, while $P_B$ introduces a call to that function using its previous signature [11][13]. Each individual modification is correct individually; as a whole, however, they generate a faulty Abstract Syntax Tree (AST).

\textbf{Conflict Layer 3. Semantic Conflict.} The last layer represents the deepest form of interference: after generating a valid textual merge with no compilation errors, there exists logical contradiction(s) between the functional execution assumptions/state-invariant produced by $P_A$ and those created by $P_B$---the Interference Problem Formalized in Classical Program Integration Theory [10][12].

This research focuses on evaluating the textual-layer. Due to their lack of knowledge of language-runtimes, environment variables or external mock database configurations textual-conflict detection provides a higher degree of both reproducibility and scalability than build- or semantic-conflict detection. Most important; since no build- or semantic-conflict can occur until a successful textual-conflict has occurred, the empirically determined conflict-rate here is conservative lower-bound of the total amount of system-wide friction.

\section{Previous Research}
Our research sits at the intersection of long-standing research in the area of repository mining and rapidly developing research in the use of large language models (LLM's) for software development.

\textbf{Human merge-conflict experience research.} Over the past twenty years, numerous researchers have investigated the ways humans experience issues involving merge conflicts in software repositories. Specifically, Ghiotto et al.\ created a definitive classification scheme of text-based collision types based upon thousands of open-source Java projects [8]. Speculative merging tools developed by Brun et al.\ were designed to reduce ``human friction'' associated with resolving merge-conflicts by identifying potential conflicts in advance via workspace analysis, separating out these conflicts prior to reaching the central origin repository [9]. These works provide foundational knowledge for understanding human experience during software maintenance. However, they make assumptions about human limitations: that developers will review code in increments, adhere to a pre-determined shift schedule, and be limited by the rate of their typing/cognitive processing abilities.

\textbf{Quantitative assessments of output quality of autonomous agents.} As coding agents continue to evolve towards greater sophistication (Devin, Claude Code, GitHub Copilot Workspace, etc.), there is increasing pressure to quantitatively assess the quality of code generated by these agents. Traditionally, most quantitative evaluations of coding agents have evaluated the accuracy of individual patches generated by coding agents, or the likelihood that individual PRs will be rejected by human maintainers [4, 5]. More recently, researchers have reported increases in both local code churn and message--code inconsistencies within code rewritten by LLMs [6, 7]. A critical void in this literature remains the assessment of interactive concurrency---how these autonomous agents interact competitively or cooperatively when deployed simultaneously across shared source code.

\textbf{Agent-based merge conflicts.} Most analogous to our current investigation, Ogenrwot and Businge created AgenticFlict---a very large-scale dataset that resulted from simulating the merge of every agentic PR against its base branch, reporting that 27.67\% of agentic PRs conflicted with their bases [3]. The authors describe their results in terms of how far a single PR has deviated from a continuously evolving trunk. Our investigation seeks to address a complementary question about collaboration: how often do two concurrently open agentic PRs conflict with one another when merged? To date we are aware of no prior work that evaluates inter-PR conflict among concurrently open agentic PRs; identifies the proportion of intra-agent vs.\ cross-agent pairs involved in such instances; or develops a taxonomy that describes what types of inter-PR conflicts exist.

\section{Evaluation Pipeline}
The evaluation pipeline utilized in this project utilizes the AIDev-pop dataset [2]; extracting a core analysis window consisting of 33,596 unique agentic pull requests mapped across 2,807 unique repositories.

\subsection{Formalizing Co-activity (RQ1--RQ2)}
A pull request can be formally expressed as follows:
\begin{equation}
P_i = (R_j,\, A_i,\, T_{\text{open}},\, T_{\text{close}})
\end{equation}
Where $R_j$ represents the specific host repository where the pull request resides; $A_i$ represents the agent model that authored the pull request; $T_{\text{open}}$ represents the point in time when the pull request was created; and $T_{\text{close}}$ represents the time at which the pull request was finally resolved (i.e., successfully merged, closed). In order for two pull requests located within the same repository to be considered co-active with respect to a given $k$-day time-padding window, they must overlap in their effective activities windows:
\begin{equation}
I_i = [\,T_{\text{open}} - k,\; T_{\text{close}} + k\,].
\end{equation}
If a PR had not yet been resolved at the time of data cut-off, then $T_{\text{close}}$ is bounded to the global data-cut off epoch. Intra-agent pairs consist of two pull requests authored by the same agent-model; cross-agent pairs are defined as all pairs for which $A_1 \neq A_2$. The interval-sweep is applied to the entirety of sample-space; and prevalence is computed for each of four different threshold values for $k$: $k \in \{0, 1, 3, 7\}$.

\subsection{Large-Scale Merge Replay (RQ3)}
We developed an automated process to find conflicts on the repositories in our corpus. This process is a merge-replay, multi-threaded, unattended. We selected a stratified sampling of pairs of co-active repositories. The reason we chose this method is that it would keep a few very highly active repositories from skewing the results. There are two strata:
\begin{itemize}
\item \textbf{Stratum A:} intra-agents, i.e., same agent, there are 625 distinct repositories. Each repository will contribute one pair of co-active.
\item \textbf{Stratum B:} Cross-Agents, i.e., different agents, since only 122 repositories out of the entire corpus show some degree of cross-agent activity (See Section~\ref{sec:rq2}), we included all of those and created one pair per repository. Thus we have a total of 747 pairs.
\end{itemize}
For each pair ($P_A$, $P_B$), the process first creates an empty bare git workspace. Then the process fetches the head commit(s) of the respective pulls over the network:
\begin{equation}
\texttt{refs/pull/N/head} \rightarrow C_A, \qquad \texttt{refs/pull/M/head} \rightarrow C_B .
\end{equation}
Then it finds the latest common ancestor ($C_{\text{base}} = \texttt{git merge-base}\; C_A\; C_B$). If it needs to, it refetches deeper because it may need to. Finally it performs a three-way merge in memory:
\begin{center}
\texttt{git merge-tree -{}-write-tree \$C\_A \$C\_B}
\end{center}
Because this path does not touch the working directory. So 0 means the merge worked fine, 1 means the merge did not work (i.e.\ conflict), and the paths of the conflicting files along with their respective git conflict type annotations are extracted from the merge's output for our taxonomy (See Section~\ref{sec:rq3}). All pairs whose commits had been force-deleted from github or whose commits had no common merge base were recorded as unavailable (i.e.\ no replacement) so that the denominators of our rates reflect what was actually available to test. We provide Wilson intervals for the rates of conflict within each stratum.

\section{Results}
\subsection{RQ1 -- Concurrency Prevalence}
The five main agents are OpenAI Codex (21,799 pull requests), Github Copilot (4,970), Devin (4,827), Cursor (1,541) and Claude Code (459). An interval sweep verifies that concurrent agentic activity is commonplace in automated environments. Of the 2807 repositories monitored by our study, 56.1\% (1574 of 2807) had at least two pull requests generated by an agent. When examining concurrency using the strictest definition based on time ($k=0$)---i.e.\ both branches are open at the exact same time---1129 of 2807 repositories have parallel workflows: 40.2\% of all repositories demonstrate this behavior (CI 95\% [38.4\%, 42.0\%]). At the pull request level, 79.4\% of all pull requests submitted by agents (26691 of 33596) are open concurrently with another pull request authored by another agent.

\begin{figure*}[t]
\centering
\includegraphics[width=\textwidth]{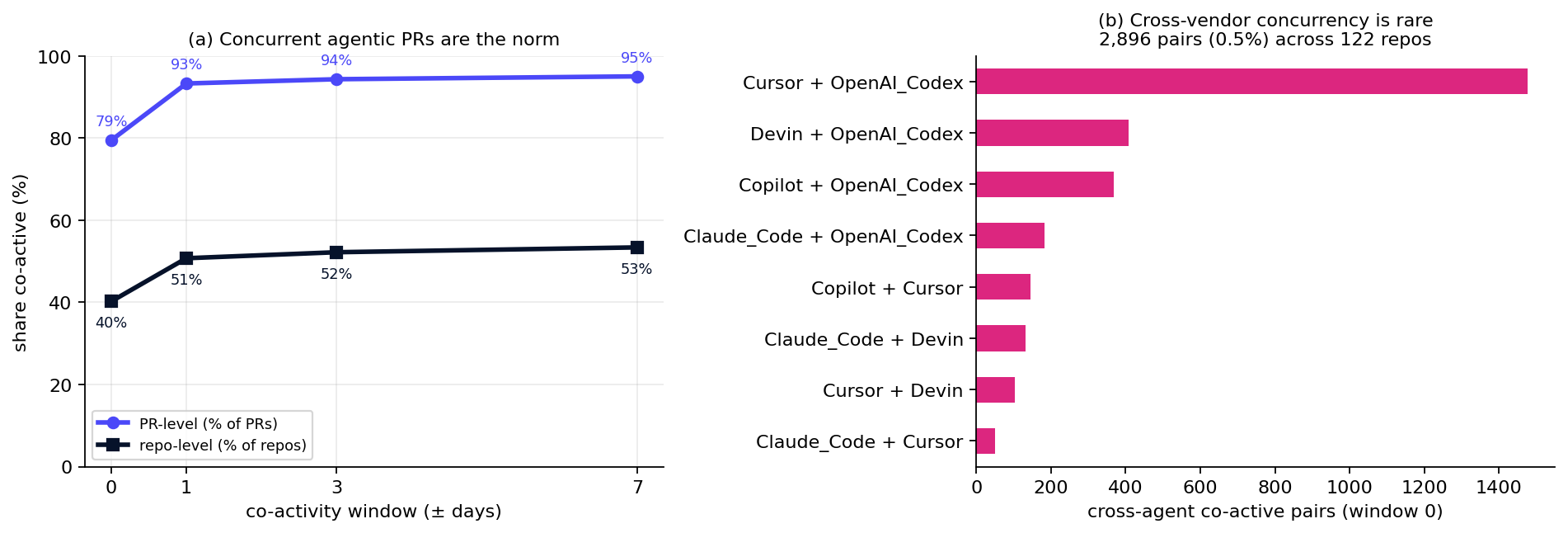}
\caption{Concurrency in the AIDev-pop corpus: (a) PR- and repository-level co-activity as the temporal-overlap window $k$ expands from 0 to $\pm 7$ days; (b) the most frequent cross-agent co-active pairings, dominated by combinations involving OpenAI Codex.}
\label{fig:coactivity}
\end{figure*}

Adding temporal padding to this definition of concurrency shows large increases in these percentages. When defining concurrency as happening in a window of $\pm 1$ day ($k=\pm 1$), the percentage of pull requests being opened contemporaneously jumps to 93.3\%. Expanding our view even further to include a window of $\pm 7$ days ($k=\pm 7$) yields virtually universal rates: nearly all (95\%) agentic pull requests happen while another agentic branch is open in approximately 53.4\% of repositories (see Table~\ref{tab:rq1} and Figure~\ref{fig:coactivity}a).

\begin{table*}[t]
\centering
\caption{Co-activity prevalence among autonomous-agent pull requests across expanding temporal-overlap windows ($k$). Derived from 33,596 PRs across 2,807 repositories.}
\label{tab:rq1}
\begin{tabular}{rrrlrlr}
\toprule
Window ($k$ days) & Total Co-active Pairs & Repos Co-active & Repo Prev.\ [95\% CI] & PRs Co-active & PR Prev.\ [95\% CI] & Cross-Agent Pairs \\
\midrule
0 & 580,913    & 1,129 & 0.402 [0.384, 0.420] & 26,691 & 0.794 [0.790, 0.799] & 2,896 (0.50\%) \\
1 & 2,755,919  & 1,424 & 0.507 [0.489, 0.526] & 31,335 & 0.933 [0.930, 0.935] & 3,690 (0.13\%) \\
3 & 6,174,411  & 1,465 & 0.522 [0.503, 0.540] & 31,683 & 0.943 [0.941, 0.945] & 4,962 (0.08\%) \\
7 & 11,926,685 & 1,498 & 0.534 [0.515, 0.552] & 31,916 & 0.950 [0.948, 0.952] & 7,681 (0.06\%) \\
\bottomrule
\end{tabular}
\end{table*}

\subsection{RQ2 -- Composition Profile}
\label{sec:rq2}
The distribution of co-active pairs demonstrates the homogeneity of contemporary agentic workflows. At the rigid baseline of $k=0$, only 2,896 of 580,913 co-active pairs represent cross-agent interactions (only 0.50\% of all concurrent pairs), distributed among only 122 of the 2,807 total repositories. Additionally, the density of cross-agent interaction is heavily weighted toward OpenAI Codex, which---by virtue of its volume---acts as the majority anchor of concurrent activity (Cursor--Codex = 1,476 pairs; Devin--Codex = 407; Copilot--Codex = 369; Figure~\ref{fig:coactivity}b). Therefore, multi-agent deployments today do not represent a heterogeneous combination of agents working in tandem; rather, the current concurrency burden comes from a single agent platform creating multiple automated PRs concurrently.

\subsection{RQ3 --- The Rates \& Taxonomies of Textual Conflicts}
\label{sec:rq3}
\begin{figure}[t]
\centering
\includegraphics[width=0.92\columnwidth]{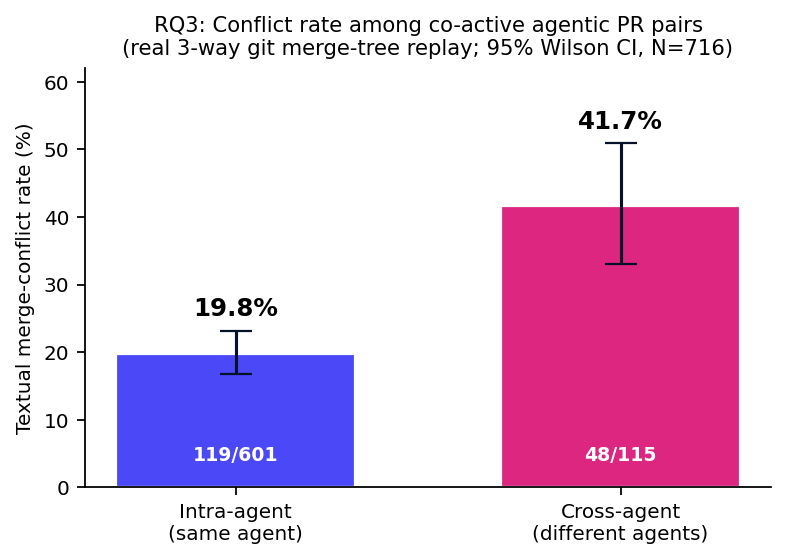}
\caption{Textual-conflict rate among co-active agentic PR pairs by stratum, from the 747-pair merge replay. Error bars are 95\% Wilson intervals ($N=716$ evaluable).}
\label{fig:rates}
\end{figure}

We merged-replayed all 747 sampled pairs (i.e., we replayed all of them) and found that 716 were evaluatable (therefore, 31 could not be evaluated because 25 had their PR reference force-deleted and 6 had no accessible merge base -- thus, 95.8\% of the possible sample set). There is substantial friction associated with the process of integrating changes across multiple agents.
\begin{itemize}
\item \textbf{Intra-Agent Pairs} (601 evaluatable out of 625): 19.8\% textual conflict rate (119/601, 95\% CI [16.8\%, 23.2\%]).
\item \textbf{Cross-Agent Pairs} (115 evaluatable out of 122): 41.7\% textual conflict rate (48/115, 95\% CI [33.1\%, 50.9\%]).
\end{itemize}
Therefore, as illustrated in Fig.~\ref{fig:rates}, cross-agent conflicts occur at approximately double the rate of intra-agent conflicts. Also, as the confidence intervals of both values do not overlap, there exists a clear distinction in the levels of conflict between these two categories. Since each repository was selected only one time during sampling, these values are free from the influence of a few extremely high activity repositories. These values represent the lack of awareness regarding concurrent PR threads among active parallel agents regardless of whether those PRs originate from the same platform or not. As intra-agent pairs comprise 99.5\% of all co-activity (see Sect.~\ref{sec:rq2}), the overall conflict rate for the population is heavily dependent upon the nearly 20\% textual conflict rate for intra-agent pairs. The above stated values only indicate conflict within the textual layer (Sect.~2) and therefore provide a lower-bound estimate of the overall inter-PR conflict level.

\begin{figure}[t]
\centering
\includegraphics[width=\columnwidth]{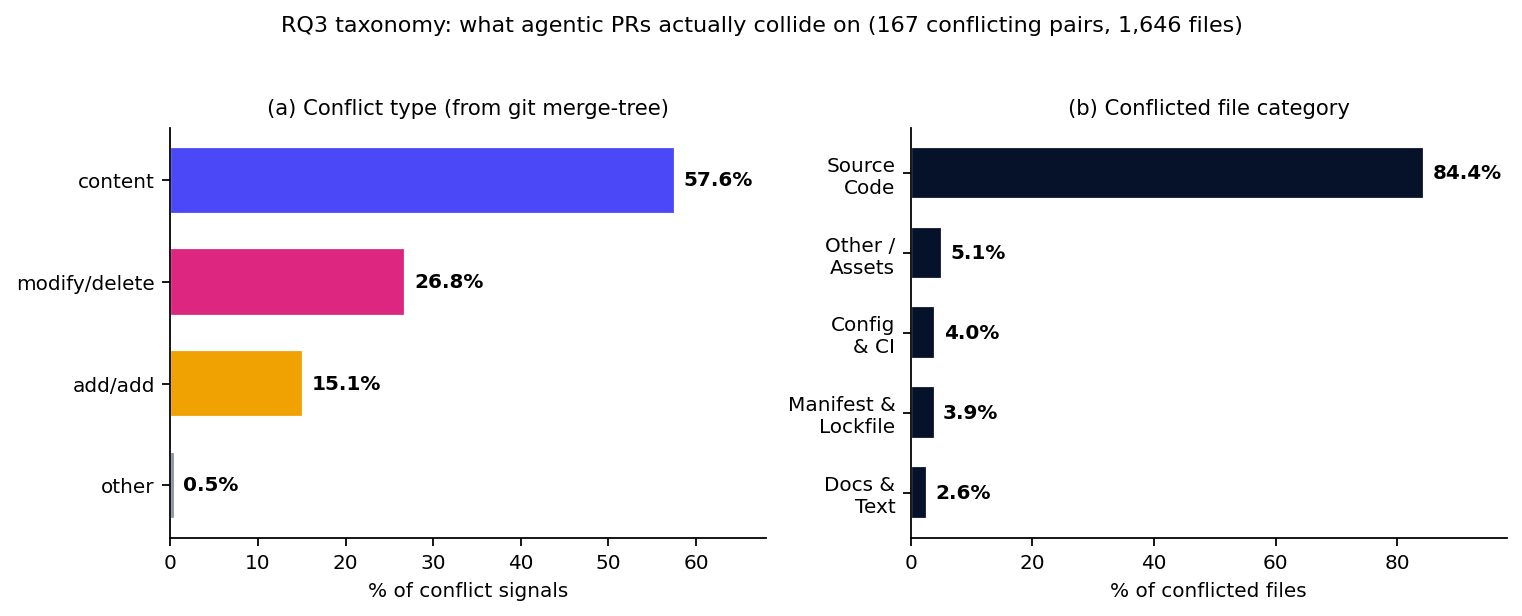}
\caption{What agentic PRs collide on: (a) conflict type from \texttt{git merge-tree}; (b) category of conflicted files. Conflicts are overwhelmingly in source code, and a large share are structural (modify/delete or add/add) rather than simple content overlaps.}
\label{fig:tax}
\end{figure}

To further illustrate how concurrently operating agents interact and cause conflicts, we analyzed the paths where conflicts occurred and the types of conflicts identified by git merge-tree for all 167 conflicting pairs (1,646 total conflicted files); See Table~\ref{tab:tax} and Fig.~\ref{fig:tax}.

\textbf{What Causes Conflict?} Contrary to the common assumption that dependency files will be involved in conflicts most frequently, conflicts fall predominantly in source code: 84.4\% of the files involved in conflicts are source files while the remaining percentage are divided equally between assets (5.1\%) configuration and CI files (4.0\%), dependency manifests and lockfiles (3.9\%) and documentation (2.6\%). When calculating by pair rather than by file, 66.5\% of the conflicting pairs contain source files while 19.2\% of the conflicting pairs include either a dependency manifest or lockfile---a real but certainly second-order phenomenon.

\textbf{How It Collides.} Using the classification defined by git as well as reporting conflict type, 57.6\% of conflict reports are content-conflicts (two different updates to the same area of a file) while a great deal of structural conflict also takes place: specifically, modify/delete conflicts account for 26.8\% (one agent has updated a file while the other has removed it) and add/add conflicts account for 15.1\% (each agent has added a version of the same file but with differing contents). This structural representation is an identifiable characteristic of un-coordinated parallel agents: they do not simply update the same lines---they often disagree as to whether a file should exist at all and/or independently create versions of the same new file.

\begin{table}[t]
\centering
\caption{Structural taxonomy of agentic merge conflicts, parsed from the \texttt{git merge-tree} output of all 167 conflicting pairs (1,646 conflicted files). Left: conflict type reported by \texttt{git}. Right: category of each conflicted file.}
\label{tab:tax}
\begin{tabular}{lr@{\hskip 2em}lr}
\toprule
\multicolumn{2}{l}{\emph{Conflict type}} & \multicolumn{2}{l}{\emph{Conflicted file category}} \\
\midrule
content       & 57.6\% & Source code          & 84.4\% \\
modify/delete & 26.8\% & Other / assets       & 5.1\% \\
add/add       & 15.1\% & Config \& CI         & 4.0\% \\
other         & 0.5\%  & Manifest \& lockfile & 3.9\% \\
              &        & Docs \& text         & 2.6\% \\
\bottomrule
\end{tabular}
\end{table}

\subsection{RQ4 -- Associations with Concurrency}
Co-activity also appears to rise as a repository's internal throughput increases. When categorizing repositories by their internal agentic volume---i.e., how many PRs occur within a repository over time---PR-level co-activity rose as well. Approximately 53\% of low-throughput repositories ($<6$ total agent PRs) exhibited PR-level co-activity; however, the rate of PR-level co-activity increased across all other bins: approximately 68\% in intermediate-throughput repositories (6--20 agent PRs); approximately 85\% in high-throughput repositories (21--100 agent PRs); and approximately 90\% in the highest-throughput repositories ($>100$ agent PRs).

As for authoring agents, the rates of PR-level co-activity vary significantly. Devin achieved the highest rates of PR-level co-activity at 82.5\%, followed by OpenAI Codex at 80.9\%, Copilot at 76.7\%, Cursor at 69.8\%, and Claude Code at 39.7\%. Differences in the rates of PR-level co-activity among different authoring-agent platforms appear to be largely influenced by the manner in which they have been deployed rather than any properties inherent to those platforms.

\begin{figure*}[t]
\centering
\includegraphics[width=\textwidth]{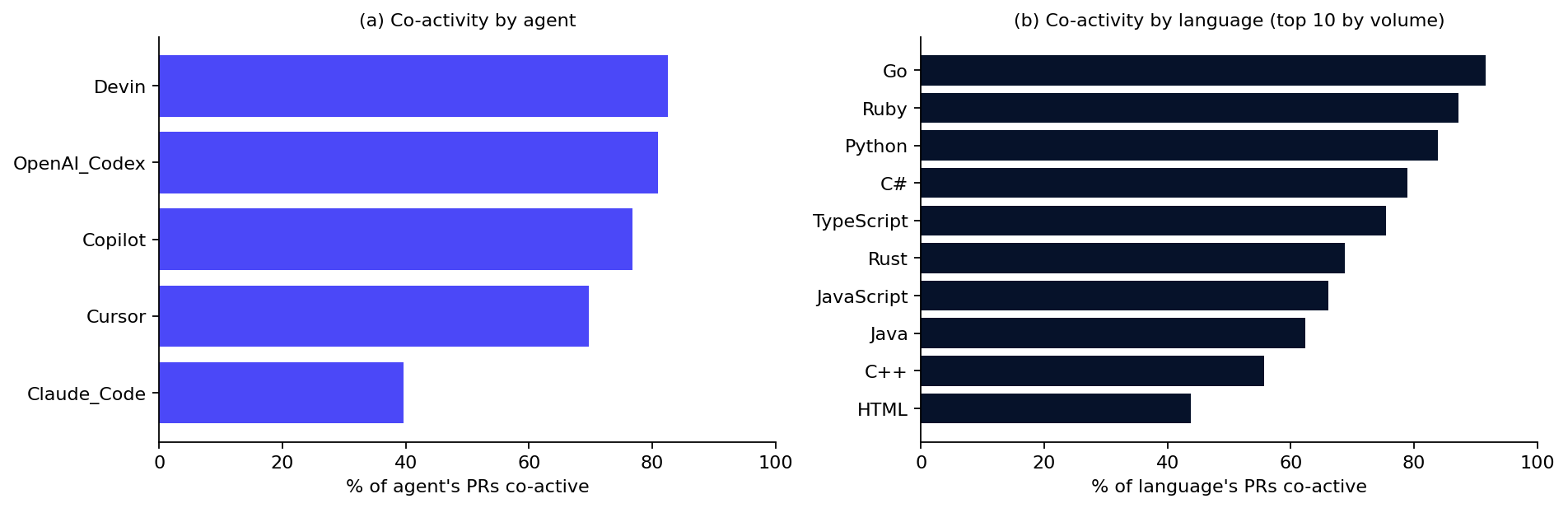}
\caption{Descriptive associations of co-activity: (a) by authoring agent; (b) by repository primary language (top 10 by PR volume). Uncontrolled trends; see the caveats in the text.}
\label{fig:assoc}
\end{figure*}

Finally, when examining whether repositories' languages influence the degree of co-activity that occurs within them, repositories utilizing Go experienced the highest rates of co-activity at an approximate 92\%, followed by repositories primarily using Ruby at approximately 87\%, Python at an approximate 84\%, and TypeScript at an approximate 75\%. Conversely, co-activity was lowest in repositories primarily written in C++ (at an approximate 56\%) and repositories primarily written in HTML (at an approximate 44\%). These relationships are observational associations only: no statistical significance testing was performed, nor were adjustments made for potential confounding variables (e.g., repository size, number of maintainers), so these figures represent descriptive trends, and the rounded percentage values reflect broad binning (see Figure~\ref{fig:assoc}).

\section{Discussion}
\subsection{The Reality of Autonomous Software Engineering}
Our findings alter how we view automated software development. Much current research on multi-agent orchestration views the concept from a theoretical perspective of an eventual future with heterogeneous, multi-vendor agent systems. In contrast, our composition data (RQ2) reveals that intra-agent concurrency is the dominant mode of composition at present. Thus, the most significant operational issue is not coordination among different vendors' tools, but rather management of the high-frequency output of a single agent framework.

Additionally, the results of our large-scale merge-replay analysis reveal that these parallel workflows create considerable integration friction. The textual conflict rate for identical agent instances was 19.8\%, indicating that currently deployed agents do not have even the most basic level of horizontal awareness. Agents operate independently and in isolation, without knowledge that other agents of the same type are simultaneously accessing and altering the same files.

Finally, the taxonomy shows that the majority of these conflicts occur at the source-code level, as opposed to the dependency-manifest and lockfile level. Further, approximately 42\% of conflict signals are structural (modify/delete or add/add operations), indicating that agents are not only editing the same lines, but also disagreeing about whether a file should exist at all.

\subsection{The Hidden Operational Costs of Agent Churn}
This lack of coordination likely results in significant additional overhead downstream. In this paper we do not measure CI/CD logs, token usage, or maintainer time; therefore, the examples provided below are hypothesized outcomes of the conflict rates that we measured, as opposed to findings based on direct observation:
\begin{enumerate}
\item \textbf{Lost CI/CD compute:} testing suites consume compute cycles validating multiple feature branches simultaneously that can never merge, due to conflicts inherent from the moment each was created in parallel.
\item \textbf{Token budget depletion:} when a conflict arises, an agent pipeline will either stall or invoke another agent instance to resolve the conflict markers, resulting in increased overall token cost.
\item \textbf{Maintainer fatigue:} human maintainers are likely required to reconcile overlapping changes to the same source files---and files modified by one agent instance and deleted by another---thereby undermining the purpose of automating engineering.
\end{enumerate}
Quantifying these potential costs is a logical subsequent research topic.

\section{Threats to Validity}
\textbf{Construct Validity.} We treat temporal co-activity as a proxy for the occurrence of concurrent work, and to prevent arbitrary threshold decisions we provide the sensitivity of co-activity from 0 through 7 days. Textual conflict is calculated using true three-way merges with \texttt{git merge-tree}---exactly how \texttt{git} reports integration---but this captures only the surface layer; no tracking is done for deeper build or semantic conflicts. Our results therefore provide a conservative lower bound.

\textbf{Internal Validity.} The total number of co-active pairs is extremely biased toward a very small subset of highly active repositories (the top ten account for 91.4\% of all raw pairs). To address this bias, we normalize prevalence at both the repository and pull-request levels and enforce one co-active pair per distinct repository within the merge-replay sample. The cross-agent stratum is limited by the amount of data available: only 122 repositories exhibited any form of cross-agent co-activity; it therefore uses a much smaller sample ($N=115$) than the intra-agent stratum ($N=601$), while 716 of the 747 pairs (95.8\%) were evaluable. Associations found in RQ4 are descriptive, uncontrolled, and should not be interpreted as causal.

\textbf{External Validity.} The findings reported here are constrained by the AIDev-pop dataset: open-source projects and five agent profiles, all over a fixed time frame (December 2024--July 2025). It is possible that the low rates of cross-agent concurrency also reflect the methods used to collect AIDev, since AIDev mined each agent independently, and because this study examines an initial period in which multi-vendor usage had only begun. These patterns may differ in proprietary enterprise code bases with stricter branching rules.

\section{Conclusion and Next Steps}
This research was the largest empirical study yet conducted into simultaneous concurrent AI-agent pull requests on GitHub. Concurrent execution is almost universally found in automated software development---it has been demonstrated by this data set that: 79.4\% of agent pull requests execute in a state of temporal overlap with other pull requests and for the majority of cases this occurs within an intra-agent (agent-specific) environment and not in cross-vendor deployments. Through the very large-scale merge replay of 747 unique pairs we have shown that when the various pull request work flows run independently of one another they generate textual merge conflicts for 19.8\% of homogeneous (pull requests executed by same-agent) scenarios and 41.7\% of cross-agent deployments; there is no overlap between the two confidence intervals. A taxonomy of actual git conflict output indicates that these conflicts occur primarily in source code (84.4\% of all files involved in a conflict are source code files vs.\ 3.9\% dependency manifests and locks); further, we have determined that 42\% of the conflicts were structural (modify/delete or add/add) whereas 58\% were content-based merges.

This demonstrates conclusively that uncoordinated reactive development paradigms will continue to fail to provide adequate support for development. Research going forward should develop programmatic orchestration mechanisms, define formalized live workspace communication protocols for active agents, and expand merge-replay pipeline methodologies to include sandboxed runtime environments such that there will exist a systematized approach to measure build failure, semantic failure and the long-term operational cost (CI compute, token spend, etc.) that has only been hypothesized thus far.

\section*{Availability of Data}
A replication package accompanies this manuscript. The package contains the pair-level merge-replay results for all 747 sampled pairs including the conflicted file path(s) and type of git conflict experienced by each conflicting pair; the per-stratum rate and taxonomy tables and the sampling, replay, and analysis scripts utilized. Each numerical value displayed in section~\ref{sec:rq3}, Table~\ref{tab:tax} and figures~\ref{fig:rates}--\ref{fig:tax} may be recalculated using the pair-level data provided in the replication package. The replication package is permanently archived at \url{https://doi.org/10.5281/zenodo.21186464} and is viewable at \url{https://github.com/Quantum535/concurrent-agentic-prs-replication}. The original PR corpus used for this research is the publicly available AIDev-pop dataset [2].

\end{document}